\documentclass[aps,prd,onecolumn,showpacs,amsmath,amssymb]{revtex4}
\usepackage{graphicx}
\usepackage{dcolumn}
\usepackage{bm}
\usepackage{amsmath}
\usepackage{amsfonts}
\usepackage{amssymb}
\def\beqra{\begin{eqnarray}}
\def\eeqra{\end{eqnarray}}
\def\beq{\begin{equation}}
\def\eeq{\end{equation}}

\def\vp{\varphi}

\def\D{\nabla^2}

\def\p{\partial}

\def\de{\delta}

\def\a{\alpha}
\def\b{\beta}
\def\n{\nu}
\def\m{\mu}
\def\k{\kappa^2}
\def\ps{\psi_1}
\def\ph{\phi_1}

\def\mh{\mathcal{H}}
\def\mr{\mathcal{R}}
\def\mo{\mathcal{O}}
\def\mp{\mathcal{P}}
\def\vn{\varphi_N}

\def\fr{\frac}
\def\pr{\prime}
\def\r{\rho}
\def\tx{\tilde{x}}

\begin{document}
\title{Unified treatment of cosmological perturbations from super-horizon 
to small scales}
\author{Carmelita Carbone}
\email{carbone@sissa.it}
\affiliation{SISSA/ISAS, Via Beirut 4, I-34014, Trieste, Italy}
\author{Sabino Matarrese}
\email{matarrese@pd.infn.it}
\affiliation{Dipartimento di Fisica `Galileo Galilei', Universit\`a di 
Padova and INFN,
Sezione di Padova, via Marzolo 8, I-35131 Padova, Italy}
\date{\today}

\begin{abstract}
We study the evolution of cosmological perturbations, using a \emph{hybrid}
approximation scheme which upgrades the \emph{weak-field} limit of Einstein's
field equations to account for \emph{post-Newtonian} scalar and vector metric
perturbations and for leading-order source terms of gravitational waves,
while including also the first and second-order perturbative approximations.
Our equations, which are derived in the Poisson gauge, provide a unified
description of matter inhomogeneities in a Universe filled with a pressureless
and irrotational fluid and a cosmological constant, ranging from the linear to
the highly non-linear regime. The derived expressions for scalar, vector
and tensor modes may have a wide range of cosmological applications, including
secondary CMB anisotropy and polarization effects, cosmographic relations
in a inhomogeneous Universe, gravitational lensing and the 
stochastic gravitational-wave backgrounds generated by non-linear cosmic 
structures.
\end{abstract}

\maketitle


\section{Introduction}

In recent years important results have been obtained in Cosmology 
from several observations of Cosmic Microwave Background (CMB) 
anisotropy and polarization, weak gravitational lensing effects and 
large-scale structure by means of galaxy redshift catalogs.  
The increasing precision that has been, and will be 
achieved by future experimental determinations requires comparable accuracy in
the theoretical estimate of the several contributions to these effects. 
Accurate study of the evolution of cosmological metric perturbations 
is therefore crucial for understanding these contributions.  
Different kinds of techniques have been developed for 
this analysis, depending on the specific range of scales under consideration. 
For example, on scales well inside the Hubble horizon, but 
still much larger than the Schwarzschild radius of collapsing bodies, 
the study of gravitational instability of collisionless
matter is performed using the Newtonian approximation. It consists of
inserting in the line-element of a Friedmann-Robertson-Walker (FRW) background
the lapse perturbation $2\varphi_N/c^2$, where $\varphi_N$ is related to the
matter density fluctuation $\delta\rho$ via the cosmological Poisson equation
$\nabla^{2}\varphi_N=4\pi G a^{2} \delta\rho$. The dynamics of the system is
then studied in Eulerian coordinates by accounting for the Newtonian mass and
momentum conservation equations, owing to the fact that the peculiar matter
flow $v$ never becomes relativistic \cite{peebles}. The Newtonian
limit, according to which the gravitational field $\varphi$ is always much
less than the square of the speed of light, $c^{2}$, can be improved by a
post-Newtonian (PN) approach to account for the moderately strong
gravitational fields generated during collapse. In this case, by considering
the expansion of the general relativistic equations in inverse powers of the 
speed of light, it is possible to neglect terms of order $(v/c)^{4}$ in the
equations of motions, i.e. to perform a first post-Newtonian (1PN)
approximation, which in Eulerian coordinates accounts for non vanishing shift
components and for an extra perturbation term $-2\varphi_N/c^{2}$ in 
the spatial
part of the line-element. Calculations using higher and higher orders of $1/c$
would generally lead to a more accurate description of the system, e.g.
accounting for the generation of gravitational waves, and possibly allow for
an extension of the range of scales to which the formalism can be applied. A
PN approach to cosmological perturbations has been followed in
Refs.~\cite{tomita3,tomita4,sa}, using Eulerian coordinates, while
Ref.~\cite{mt}, uses Lagrangian coordinates. 

On the other hand, the first-order perturbations for non-relativistic matter,
obtained with the Newtonian treatment coincide with the results of linear
general relativistic perturbation theory in the so-called longitudinal gauge
\cite{berts}. The relativistic linear perturbative approach is the one applied
to the study of the small inhomogeneities giving rise to large-scale
temperature anisotropies of the CMB. However, on small and intermediate
scales, linear theory is no longer accurate and a general fully relativistic
second-order perturbative technique is required. In fact, second-order metric
perturbations determine new contributions to the CMB temperature anisotropy
\cite{pyne,mm} and polarization \cite{mhm}. In particular, second-order
scalar, vector and tensor metric perturbations produce secondary anisotropies
in the temperature and polarization of the CMB which are in
competition with other non-linear effects, such as that due to weak
gravitational lensing produced by matter inhomogeneities, which induces the
transformation of E-mode into B-mode polarization \cite{sz}.

Moreover, accounting for second-order effects helps to follow the
gravitational instability on a longer time scale and to include new non-linear
and non-local phenomena. The pioneering work in this field is due to Tomita
\cite{tomita} who performed a synchronous-gauge calculation of the
second-order terms produced by the mildly non-linear evolution of scalar
perturbations in the Einstein-de Sitter Universe. An equivalent result, was
obtained with a different technique in Ref.~\cite{mps}. The inclusion of
vectors and tensor modes at the linear level, acting as further seeds for the
origin of second-order perturbations of any kind was first considered in
Ref.~\cite{tomita2}; in Ref.~\cite{mmb} the evolution of relativistic
perturbations in the Einstein-de Sitter cosmological model was considered and
second-order effects were included in two different settings: the synchronous
and the Poisson gauge.

As we have stressed, the evolution of cosmological perturbations away from the
linear level is rich of several effects as, in particular, mode-mixing which
not only implies that different Fourier modes influence each other, but also
that density perturbations act as a source for curl vector modes and
gravitational waves.\newline

The aim of the present paper is therefore to obtain a unified treatment able
to follow the evolution of cosmological perturbations from the linear to the
highly non-linear regime. As we will show hereafter, this goal is indeed
possible on scales much larger than the Schwarzschild radius of collapsing
bodies, by means of a ``hybrid approximation'' of Einstein's field equations,
which mixes post-Newtonian (PN) and second-order perturbative techniques to
deal with the perturbations of matter and geometry. In our study we adopt the
Poisson gauge \cite{maber}, which, being the closest to the Eulerian Newtonian
gauge, allows a simple physical interpretation of the various perturbation
modes. We derive a new set of equations which holds on all cosmologically
relevant scales and allows to describe matter inhomogeneities during all the
different stages of their evolution. The new approach gives a more
accurate description of the metric perturbations generated by non-linear
structures than the second-order perturbation theory, which can only account
for small deviations from the linear regime. For example, on small scales our
set of equations can be used to provide a PN description of metric
perturbations generated by highly non-linear structures within dark matter
halos, while describing their motion by means of the standard Newtonian
hydrodynamical equations. On large scales our equations converge to the first
and second-order perturbative equations as obtained in Ref.~\cite{mmb} 
(see also Refs.~\cite{acqua,noh}), which
implies that they can be applied to every kind of cosmological sources.\newline

The plan of the paper is as follows. In Section II we obtain metric
perturbations in the Poisson gauge according to our hybrid approximation. In
Section III we obtain the source terms for scalar, vector and tensor metric
perturbations and the evolution equations for the matter density and peculiar
velocity. Section IV is devoted to a comparison of our approach with known
approximation schemes, such as the standard Newtonian one, the linear and the
second-order relativistic perturbative approaches, in the appropriate regimes
of applicability. Moreover, we derive a PN approximation to describe the
highly non-linear regime on small scales. 
In Section V we show that the PN expressions for vector and tensor modes  
actually hold on all cosmological scales. In Section VI we sketch some
cosmological applications of our approach and we draw our main conclusions. An
Appendix is devoted to the solution of the inhomogeneous gravitational-wave 
equation.

\section{Scalar, vector and tensor metric perturbation modes}

Adopting the conformal time $\eta$, the perturbed line-element around a
spatially flat FRW background in the Poisson gauge \cite{berts,mmb} 
takes the form
\begin{align}
\label{gij}
ds^{2} = a^{2}(\eta) \left[  -(1+2\phi) d \eta^{2} - 2 V_{\alpha}
d\eta dx^{\alpha}+ \left( (1-2\psi)\,\delta_{\alpha\beta}+h_{\alpha\beta}
\right)  dx^{\a}dx^{\b}\right]  \;.
\end{align}
In Eq.~(\ref{gij}) the metric includes perturbative terms of any order around
the FRW background. In this gauge, $V_{\alpha}$ are pure vectors, i.e. they
have vanishing spatial divergence, $\partial^{\alpha}V_{\alpha}=0$, while
$h_{\alpha\beta}$ are traceless and transverse, i.e. pure tensor modes,
$h^{\alpha}{}_{\alpha}= \partial^{\alpha}h_{\alpha\beta} =0$. Here and in what
follows spatial indices are raised by the Kronecker symbol $\delta^{\alpha}
{}_{\beta}$. Unless otherwise stated, we use units with $c=1$. From
the results of the post-Newtonian theory \cite{ch,ch2,ch3,mt}, 
we deduce that vector
and tensor metric modes, to the leading order in powers of $1/c$, are
respectively of $\mathcal{O}(1/c^{3})$ and $\mathcal{O} (1/c^{4})$, since to
the lowest order it is well known that the line-element (\ref{gij}) assumes
the weak-field form $ds^{2}=a^{2} [-(1+2\phi_{1})\,d\eta^{2}+(1-2\psi_{1})\,
\delta_{\alpha\beta}dx^{\alpha}dx^{\beta}]$, where the scalars $\phi_{1}$ and
$\psi_{1}$ are both of $\mathcal{O}(1/c^{2})$ and $\phi_{1}=\psi_{1}
\equiv\varphi$ \cite{peebles}.

Let us now write Einstein's equations $G^{i}{}_{j} = \kappa^{2} T^{i} {}_{j}$
in the perturbed form
\begin{align}
\label{EQ}
^{(0)} G^{i}{}_{j} + \delta G^{i}{}_{j} = \kappa^{2} \Big(\,^{(0)}
T^{i}{}_{j} +\delta T^{i}{}_{j}\,\Big)\;,
\end{align}
where $\kappa^{2} = 8\pi G/c^{4}$ and $^{(0)}G^{i}{}_{j}=\kappa^{2}\,
^{(0)}T^{i}{}_{j}$ reduce to the background Friedmann equations. Hereafter, we
assume that the Universe is filled with a cosmological constant $\Lambda$ and
a pressureless fluid -- made of Cold Dark Matter (CDM) plus
baryons -- whose stress energy-tensor reads $T^{i}{}_{j}=\rho u^{i} u_{j}$
($u^{i} u_{j} =-1$). In this case the Friedmann equations read 
$3\mathcal{H}^{2}= a^{2} \left(  8\pi G \bar\rho+ \Lambda\right)$ 
and $\bar\rho\,^{\prime
}= -3\mathcal{H} \bar\rho$, where primes indicate differentiation with respect
to $\eta$, $\mathcal{H} \equiv a^{\prime}/a$ and $\bar\rho$ is the mean matter
energy density. 

Since the metric (\ref{gij}) can be expressed in an
invariant form both in the PN and in the second-order perturbative
approximations, we introduce a hybrid formalism that consists in evaluating
Einstein's field equations up to the correct order in powers of $1/c$, while
including some ``hybrid'' correction terms, which cannot be completely
absorbed in a ``rigid'' PN approximation, but are required for consistency
with a second-order general relativistic approximation.

Namely, writing $\psi=\psi_{1}+\psi_{2}$ and $\phi=\phi_{1}+\phi_{2}$ (where
$\psi_{2}$ and $\phi_{2}$ in principle contain all powers of $1/c$), and
replacing the metric (\ref{gij}) in Eq.~(\ref{EQ}), the ($0$-$0$) and
($0$-$\alpha$) components of the perturbed Einstein equations take the form
\begin{align}
\label{G00}
\delta G^{0}{}_{0}= -\frac{1}{a^{2}}\,\left[  -6\mathcal{H}
\,(\mathcal{H}\phi_{1}+ \psi^{\prime}_{1}) + 2\nabla^{2}\psi+3\partial^{\nu
}\psi_{1} \partial_{\nu}\psi_{1}+ 8\psi_{1}\nabla^{2}\psi_{1}+\mathcal{S}
_{1}\right]  =\kappa^{2}\delta T^{0}{}_{0} \;,
\end{align}
where $\mathcal{S}_{1} \equiv12\mathcal{H}^{2}\phi_{1}^{2}+3\psi_{1}^{\prime
2}+ 12\mathcal{H}\phi_{1}\psi_{1}^{\prime} - 12\mathcal{H}\psi_{1}\psi_{1}
^{\prime}$, and
\begin{align}
\label{G0a}
\delta G^{0}{}_{\alpha}=-\frac{2}{a^{2}}\,\Big(\mathcal{H}
\partial_{\alpha}\phi_{1}+ \partial_{\alpha}\psi_{1}^{\prime}+\frac{1}
{4}\nabla^{2} V_{\alpha} +\mathcal{S}_{2 \alpha} \Big)=\kappa^{2}\,\delta
T^{0}{}_{\alpha}\,,
\end{align}
where $\mathcal{S}_{2 \alpha} \equiv- 4\mathcal{H}\phi_{1}\partial_{\alpha
}\phi_{1}-2\phi_{1}\, \partial_{\alpha}\psi_{1}^{\prime}- \psi_{1}^{\prime
}\partial_{\alpha}\,\phi_{1}+ 2\psi_{1}^{\prime}\partial_{\alpha}\psi
_{1}+2\psi_{1}\partial_{\alpha} \psi_{1}^{\prime}$.

The traceless part
of the ($\alpha$-$\beta$) Einstein's equations $\delta G^{\alpha}{}_{\beta
}=\kappa^{2}\,\delta T^{\alpha}{}_{\beta}$ reads
\begin{align}
\label{Gab} 
&  \left[  \frac{2}{3}\,\nabla^{2}(\phi-\psi)-\frac{8}{3}
\,\psi_{1}\nabla^{2}\psi_{1}- \frac{4}{3}\,\phi_{1}\nabla^{2}\phi_{1}+\frac
{4}{3}\,\psi_{1}\nabla^{2}\phi_{1}- \frac{2}{3}\,\partial^{\nu}\phi
_{1}\partial_{\nu}\phi_{1}+ \frac{4}{3}\, \partial_{\nu}\phi_{1}\partial^{\nu
}\psi_{1}-2\partial^{\nu}\psi_{1}\partial_{\nu} \psi_{1}\right]
\delta^{\alpha} {}_{\beta}\nonumber\\
&  -2\partial^{\alpha}\partial_{\beta}(\phi-\psi)+8\psi_{1}\partial^{\alpha
}\partial_{\beta}\psi_{1}+6\partial^{\alpha}\psi_{1}\partial_{\beta} \psi_{1}
-2\partial^{\alpha}\phi_{1}\partial_{\beta}\psi_{1}+2\partial^{\alpha}\phi_{1}
\partial_{\beta}\phi_{1}-2\partial^{\alpha} \psi_{1} \partial_{\beta}\phi_{1}
+4\phi_{1}\partial^{\alpha}\partial_{\beta} \phi_{1}\nonumber\\
&  -4\psi_{1}\partial^{\alpha}\partial_{\beta}\phi_{1}+\partial^{\alpha}
(2\mathcal{H}\,V_{\beta}+V^{\prime}_{\beta})+\partial_{\beta} (2\mathcal{H}
\,V^{\alpha} + V^{\prime\,\alpha})+h^{\prime\prime\,\alpha}{}_{\beta
}+2\mathcal{H}h^{\prime\,\alpha}{}_{\beta} -\nabla^{2} h^{\alpha}{}_{\beta
}=2a^{2}\kappa^{2}\Big(\delta T^{\alpha}{}_{\beta}-\frac{1}{3}\delta T^{\nu}
{}_{\nu} \delta^{\alpha}{}_{\beta}\Big)\,,
\end{align}
while its trace becomes
\begin{align}
\label{Gaa} 
& 2\nabla^{2}(\phi-\psi)+6\mathcal{H} \phi_{1}^{\pr}
+6(\mathcal{H}^{2}+2\mathcal{H}^{\pr})\phi_{1}+6\psi_{1}^{\prime\prime}
+12\mathcal{H}\psi_{1}^{\pr}-2\partial^{\n}\phi_{1}\partial_{\n}\phi
_{1}\nonumber\\
& -4\phi_{1}\nabla^{2}\phi_{1}-3\partial_{\n}\psi_{1}\partial^{\n}\psi
_{1}-8\psi_{1}\nabla^{2}\psi_{1}+4\psi_{1}\nabla^{2}\phi_{1}-2\partial_{\n
}\phi_{1}\partial^{\n}\psi_{1} =a^{2}\kappa^{2}\delta T^{\n}{}_{\n}\;.
\end{align}
The components of the perturbed stress-energy tensor $\delta T^{i}{}_{j}$ will
be calculated later to the correct order in powers of $1/c$. Note that Eqs.~
(\ref{G00}), (\ref{Gab}) and (\ref{Gaa}) are evaluated up to $\mathcal{O}
(1/c^{4})$, while Eq.~(\ref{G0a}) is evaluated up to $\mathcal{O}(1/c^{3})$.
However, the terms $\mathcal{S}_{1}$ and $\mathcal{S}_{2 \alpha}$ are at least
of $\mathcal{O}(1/c^{6})$ and $\mathcal{O}(1/c^{5})$, respectively, and come
out from our hybrid scheme, which mixes PN and second-order perturbative
approaches. Moreover, since our purpose is to calculate the source of
gravitational waves to the leading order in powers of $1/c$, in 
Eq.~(\ref{Gab}) we do not take into account contributions of order higher than
$\mo(1/c^4)$, though retaining time derivatives of $h^\a{}_\b$.   

Taking the divergence of Eq.~(\ref{G0a}), to solve for the
combination $\mathcal{H}\phi_{1} +\psi_{1}^{\prime}$, and replacing it in
Eq.~(\ref{G00}), we obtain
\begin{align}
\label{IG0a}
\nabla^{2}(\mathcal{H}\phi_{1}+\psi_{1}^{\prime})=-\frac{a^{2}
\kappa^{2}}{2}\partial^{\nu}\delta T^{0}{}_{\nu} -\partial^{\nu}
\mathcal{S}_{2\nu}\;,
\end{align}
and
\begin{align}
\label{psi}
\nabla^{2}\nabla^{2}\psi=\nabla^{2}\nabla^{2}(\psi_{1}+\psi_{2}) =
-\frac{a^{2}\kappa^{2}}{2} \Big[\nabla^{2}\delta T^{0}{}_{0} + 3\mathcal{H}
(\partial^{\nu} \delta T^{0}{}_{\nu})\Big] -3\mathcal{H}\partial^{\nu}
\mathcal{S}_{2 \nu}-\frac{1}{2}\nabla^{2}\mathcal{S}_{1} -\nabla^{2}
\left(\frac{3}{2}\partial^{\n}\psi_{1}\partial_{\n}\psi_{1}
+4\psi_{1}\nabla^{2}\psi_{1}\right)\,.
\end{align}
The pure vector part $V_{\alpha}$ can be isolated by replacing $\nabla
^{2}(\mathcal{H}\phi_{1}+\psi_{1}^{\prime})$ in Eq.~(\ref{G0a}), where we now
neglect the term $\mathcal{S}_{2 \alpha}$ since it is at least of $\mo(1/c^5)$
\begin{align}
\label{Va}
\nabla^{2}\nabla^{2} V_{\alpha}=2a^{2}\kappa^{2}\left(
\partial_{\alpha} \partial^{\nu}\delta T^{0}{}_{\nu}-\nabla^{2}\delta T^{0}
{}_{\alpha}\right) \;.
\end{align}
Finally, applying the operator $\partial^{\beta}\partial_{\alpha}$ to
Eq.~(\ref{Gab}), we can solve for the combination $\phi-\psi$ and write
\begin{align}
\label{phi-psi}
\nabla^{2}\nabla^{2} (\phi-\psi)  & =-\frac{3}{2}a^{2}
\kappa^{2} \partial^{\beta}\partial_{\alpha}\left(  \delta T^{\alpha}{}
_{\beta}-\frac{1}{3}\delta T^{\nu}{}_{\nu}\delta^{\alpha}{}_{\beta}\right)
+\frac{3} {2}\partial_{\alpha}\partial^{\beta}\left(  \partial_{\beta}\psi_{1}
\partial^{\alpha}\phi_{1}-\partial_{\beta}\phi_{1} \partial^{\alpha}\psi
_{1}-\partial_{\beta}\psi_{1} \partial^{\alpha}\psi_{1}-\partial_{\beta}
\phi_{1} \partial^{\alpha}\phi_{1}\right) \nonumber\\
& +\nabla^{2} \left(  \frac{9}{2}\partial_{\nu}\psi_{1} \partial^{\nu}
\psi_{1}-2\partial_{\nu}\psi_{1} \partial^{\nu}\phi_{1}+\frac{5}{2}
\partial_{\nu}\phi_{1} \partial^{\nu}\phi_{1}+4\psi_{1}\nabla^{2}\psi_{1}
+2\phi_{1}\nabla^{2}\phi_{1}-2\psi_{1}\nabla^{2}\phi_{1}\right) \;.
\end{align}
Replacing the latter expression in Eq.~(\ref{Gab}), together with the
expression for the vector mode $V_{\alpha}$ 
obtained by taking the divergence of Eq.~(\ref{Gab}), we find
\begin{align}
\label{hab}
\nabla^{2} \nabla^{2} \left(h^{\prime\prime\,\alpha}{}_{\b}
+2\mh h^{\prime\,\alpha}{}_{\b}-\D h^\a{}_\b\right)=2\k
a^2&\bigl[\D\bigl(\D\mr^\a{}_\b-\p^\a\p_\n\mr^\n{}_\b
-\p_\b\p^\n\mr^\a{}_\n \bigr) \nonumber\\
&+\fr{1}{2}\left(\D\p^\m\p_\n\mr^\n{}_\m\,\de^\a{}_\b
+\p^\a\p_\b\p^\m\p_\n\mr^\n{}_\m \right) \bigr]\;, 
\end{align}
where we have defined the traceless tensor
\begin{align}
\label{Rab}
\mr^\a{}_\b&=\de T^\a{}_\b-\fr{1}{3}\de T^\n{}_\n \de^\a{}_\b 
-\fr{1}{\k a^2}\left(\fr{2}{3}\ps\D\ph+\fr{2}{3}\p_\n\ph\p^\n\ps
-\fr{4}{3}\ps\D\ps-\fr{2}{3}\ph\D\ph-\fr{1}{3}\p^\n\ph\p_\n\ph
-\p^\n\ps\p_\n\ps\right)\de^\a{}_\b \nonumber\\
&-\fr{1}{\k a^2}\left(4\ps\p^\a\p_\b\ps+3\p^\a\ps\p_\b\ps-\p^\a\ph\p_\b\ps
+\p^\a\ph\p_\b\ph-\p^\a\ps\p_\b\ph+2\ph\p^\a\p_\b\ph
-2\ps\p^\a\p_\b\ph\right)\;.
\end{align} 
The form of Eq.~(\ref{Va}) and Eq.~(\ref{hab}) allows to directly check that
vector sources are transverse while tensor sources are doubly 
transverse and traceless.

Actually, there is a very simple way of solving the perturbed Einstein
equations $\de G^0{}_\a=\k a^2\de T^0{}_\a$ and 
$\de G^\a{}_\b=\k a^2\de T^\a{}_\b$ with respect to vectors 
$V_\a$ and tensors $h^\a{}_\b$, respectively. In fact, after retaining only 
the metric terms which appear linearly on the LHS of Eq.~(\ref{G0a}) and 
Eq.~(\ref{Gab}) \footnote{In this context we are using the traceless part of 
$\de G^\a{}_\b=\k a^2\de T^\a{}_\b$, but this is not 
strictly necessary; it is indeed sufficient to apply the projection 
operator to the whole equation $\de G^\a{}_\b=\k a^2\de T^\a{}_\b$.} 
and defining as $\mr^\a{}_\b$ the consequently obtained RHS,
it is possible to apply to both  
sides the correct combinations of the direction-independent projection
operator \cite{grav}
\begin{align}
\label{Pab}
\mp^\a{}_\b=\de^\a{}_\b-\left(\D \right)^{-1} \p^\a\p_\b
\end{align}
and automatically obtain the vectors
\begin{align}
\label{PVa}
\D V_\a=-2a^2\k \mp^\n{}_\a\left(\de T^0{}_\n
-\fr{2}{a^2\k}\mathcal{S}_{2 \n}\right) \;,
\end{align}
and tensors
\begin{align}
\label{Phab}
h^{\prime\prime\,\alpha}{}_{\b}
+2\mh h^{\prime\,\alpha}{}_{\b}-\D h^\a{}_\b=2\k
a^2\left(\mp^\a{}_\n \mp^\m{}_\b 
-\fr{1}{2}\mp^\a{}_\b \mp^\m{}_\n\right)\mr^\n{}_\m \;. 
\end{align}
After applying twice the Laplacian operator to Eq.~(\ref{Phab}) and 
neglecting as before the term $\mathcal{S}_{2 \n}$ in Eq.~(\ref{PVa}), 
we recover Eq.~(\ref{Va}) and Eq.~(\ref{hab}).

\section{The stress-energy tensor and the source of metric perturbations}

For the purpose of calculating the components of the perturbed stress-energy
tensor $\delta T^{i}{}_{j}$ to the correct order in powers of $1/c$, it is
convenient to restore the speed of light $c$ in the time coordinate
$dx^{0}=cd\eta$. From Eq.~(\ref{gij}) we obtain the four-velocity $u^{i}\equiv
dx^{i}/ds$
\begin{align}
\label{Uo}
u^{0}\simeq\frac{1}{a}\,\biggl[1-\frac{1}{2}\,\bigg(2\phi_{1}- 
\frac{v^{2}}{c^{2}}\bigg)\biggr]+\mathcal{O}\left( \frac{1}{c^{4}}\right)  \,,
\end{align}
\begin{align}
\label{Ua}
u^{\alpha}=\frac{v^{\alpha}}{c}u^{0}+\mathcal{O}\left( \frac
{1}{c^{5}}\right)  \;,
\end{align}
where $v^{2}=v^{\nu}v_{\nu}$ and $v^{\alpha}\equiv dx^{\alpha}/d\eta$ is the
coordinate three-velocity with respect to the FRW background.

The total energy-momentum tensor for our 
fluid of dust and cosmological constant reads
\begin{align}
\label{T}
T^{i}{}_{k}=\,^{(0)}T^{i}{}_{k}+\delta T^{i}{}_{k}=\left[
(\rho_{\Lambda} +\rho)c^{2}+ p_{b}\right] g_{kj}u^{i}u^{j}+p_{b}\delta^{i}
{}_{k}\,,
\end{align}
where $^{(0)}T^{i}{}_{k}$ is the background stress-energy tensor and
$\rho=\bar{\rho}+ \delta\rho$ is the total mass-density. The background
density $\rho_{b}=\bar{\rho}+\rho_{\Lambda}$ includes the contribution from
the cosmological constant, $\rho_{\Lambda}=(\Lambda c^{2})/(8\pi G)$, while
the background pressure $p_{b}=p_{\Lambda} =-\rho_{\Lambda}c^{2}$ is only due
to the latter. 

Turning to the components of the perturbed
energy-momentum tensor, in terms of the coefficients $\phi_{1}$and $\psi_{1}$
of the metric (\ref{gij}) and up to the correct orders in powers of $1/c$, we
find
\begin{align}
\label{T00}
\delta T^{0}{}_{0}=T^{0}{}_{0}-\,^{(0)}T^{0}{}_{0}=-c^{2}\delta
\rho-v^{2}\rho+ \mathcal{O} \left(  \frac{1}{c^{2}}\right)  \,,
\end{align}
\begin{align}
\label{T0a}
\delta T^{0}{}_{\alpha}=T^{0}{}_{\alpha}-\,^{(0)}T^{0}{}_{\alpha}=
v_{\alpha}\rho c\left( 1-2\phi_{1}-2\psi_{1}\right) + \frac{v_{\alpha}}{c}\rho
v^{2}+\mathcal{O} \left(  \frac{1}{c^{3}}\right)  \,,
\end{align}
\begin{align}
\label{Ta0}
\delta T^{\a}{}_{0}=T^{\a}{}_{0}-\,^{(0)}T^{\a}{}_{0}=-v^{\a}\rho
c-\frac{v^{\alpha}}{c}\rho v^{2}+\mathcal{O} \left( \frac{1}{c^{3}}\right)
\,,
\end{align}
\begin{align}
\label{Tab}
\delta T^{\alpha}{}_{\beta}=T^{\alpha}{}_{\beta}-\,^{(0)}T^{\alpha
}{}_{\beta} =v^{\alpha}v_{\beta}\rho\left( 1-2\phi_{1}-2\psi_{1}+\frac{v^{2}
}{c^{2}} \right)  + \mathcal{O}\left(  \frac{1}{c^{4}}\right)  \,.
\end{align}
In the hybrid equations which we are about to derive we will keep step by step
only the $\delta T^{i}{}_{j}$ components we need to let our set of equations
hold in the first, second perturbative order and PN regimes. 

By substituting $\delta T^{\a}{}_{\b}$ in Eq.~(\ref{phi-psi}), we obtain up to
$\mathcal{O}(1/c^{2})$
\begin{align}
\label{weak-field}
\nabla^{2} \nabla^{2} (\phi_{1}-\psi_{1})=0\;,
\end{align}
and we can safely assume $\phi_{1}=\psi_{1}\equiv\varphi$. 

This allows
to further simplify Eqs.~(\ref{Gaa})-(\ref{phi-psi}) and obtain our final set
of hybrid equations for cosmological perturbations, namely
\begin{align}
\label{IG0anew}
\nabla^{2}(\mathcal{H}\varphi+\varphi^{\prime})=-\frac{a^{2}
\kappa^{2}c^{2}}{2}\partial^{\nu}(\rho v_{\n})\;,
\end{align}
\begin{align}
\label{psinew}
\nabla^{2} \nabla^{2}\psi & =-\frac{a^{2} \kappa^{2}}{2}
\Bigl[3\mathcal{H}\partial^{\n}(\rho\, v_{\n}(1-4\varphi))-\nabla^{2}
(c^{2}\delta\rho+\rho\,v^{2})\Bigr] +3\frac{\mathcal{H}}{c}\partial^{\nu
}\left( 4\frac{\mathcal{H}}{c}\varphi\partial_{\nu}\varphi-\frac{1}{c}
\varphi^{\prime}\partial_{\nu}\varphi\right) \nonumber\\
& -\nabla^{2}\left( \frac{3}{2}\partial^{\nu}\varphi\partial_{\nu}
\varphi+4\varphi\nabla^{2}\varphi+6\frac{\mathcal{H}^{2}}{c^{2}}\varphi
^{2}+\frac{3}{2 c^{2}}\varphi^{\prime\,2}\right) \;,
\end{align}
\begin{align}
\label{vnew}
\nabla^{2}\nabla^{2} V_{\a}=2 a^{2} \kappa^{2} \left[ \partial
_{\a}\partial^{\n}(c\rho\, v_{\n})-\nabla^{2}(c\rho\, v_{\a})\right]
\end{align}
\begin{align}
\label{phi2-psi2}
\nabla^{2}\nabla^{2} (\phi-\psi) =-\frac{3}{2}a^{2}
\kappa^{2} \partial^{\mu}\partial_{\nu}\left( \rho\, v^{\n}v_{\m}-\frac{1}%
{3}\rho\, v^{2}\delta^{\nu}{}_{\mu}\right) +\nabla^{2} \left( 5 \partial_{\nu
}\varphi\partial^{\nu}\varphi+4\varphi\nabla^{2}\varphi\right)  -3\partial
_{\nu}\partial^{\mu}\left( \partial_{\mu}\varphi\partial^{\nu}\varphi\right)
\;,
\end{align}
\begin{align}
\label{phinew}
\nabla^{2} \nabla^{2}\phi & =-\frac{a^{2} \kappa^{2}}{2}
\Bigl[3\mathcal{H}\partial^{\nu}\left(\rho\,v_{\n}(1-4\varphi)\right)
+3\partial^{\mu}\partial_{\nu}(\rho\,v^{\n}v_{\m})-\nabla^{2} (c^{2}
\delta\rho+2\rho\,v^{2})\Bigr] +\frac{7}{2}\nabla^{2}(\partial_{\nu}
\varphi\partial^{\nu}\varphi)\nonumber\\
& -3\partial_{\nu}\partial^{\mu}(\partial_{\mu}\varphi\partial^{\nu}\varphi)
+3\frac{\mathcal{H}}{c}\partial^{\nu}\left( 4\frac{\mathcal{H}}{c}
\varphi\partial_{\nu}\varphi-\frac{1}{c}\varphi^{\prime}\partial_{\nu}
\varphi\right)  -\nabla^{2}\left( 6\frac{\mathcal{H}^{2}}{c^{2}}\varphi
^{2}+\frac{3}{2 c^{2}}\varphi^{\prime\,2}\right) \;,
\end{align}
\begin{align}
\label{habnew}
\nabla^{2} \nabla^{2} \left(\frac{1}{c^{2}}h^{\prime\prime\,\alpha}{}_{\b}
+\fr{2\mh}{c^2}h^{\prime\,\alpha}{}_{\b}-\D h^\a{}_\b\right)=2\k
a^2&\bigl[\D\bigl(\D\mr^\a{}_\b-\p^\a\p_\n\mr^\n{}_\b
-\p_\b\p^\n\mr^\a{}_\n \bigr) \nonumber\\
&+\fr{1}{2}\left(\D\p^\m\p_\n\mr^\n{}_\m\,\de^\a{}_\b
+\p^\a\p_\b\p^\m\p_\n\mr^\n{}_\m \right) \bigr]\;, 
\end{align} 
where the traceless tensor $\mr^\a{}_\b$, Eq.~(\ref{Rab}), now reads
\begin{align}
\label{Rabnew}
\mr^\a{}_\b=\r\left(v^\a v_\b-\fr{1}{3} v^2 \,\de^\a{}_\b \right)
-\fr{2}{\k a^2}\left(\p^\a\vp\, \p_\b\vp
-\fr{1}{3}\p^\n\vp\,\p_\n\vp\,\de^\a{}_\b \right)
-\fr{4}{\k a^2}\left(\vp\,\p^\a\p_\b\vp
-\fr{1}{3}\vp\,\D\vp\,\de^\a{}_\b\right)\;,
\end{align}
while the trace part of the ($\alpha$-$\beta$) component of Eq.~(\ref{EQ})
becomes
\begin{align}
\label{trace}
2\nabla^{2}(\phi-\psi)+18\frac{\mathcal{H}}{c^{2}}\varphi^{\pr
}+6\left( 2\frac{\mathcal{H}^{\pr}}{c^{2}} +\frac{\mathcal{H}^{2}}{c^{2}
}\right) \varphi+\frac{6}{c^{2}}\varphi^{\prime\prime}-7\partial_{\n}
\varphi\partial^{\n}\varphi-8\varphi\nabla^{2}\varphi=
\k a^2 \r \,v^2 \;.
\end{align}

Using Eqs.~(\ref{T00})-(\ref{Tab}), and the expression $\phi_{1}=\psi
_{1}=\varphi$, we can write the stress-energy tensor conservation equations
$T^{j}_{i;j}=0$ in a form that will give us the equations for our pressureless
fluid in the first, second-order and PN regimes, respectively. 

More specifically, the energy conservation equation reads
\begin{align}
\label{energy}
\rho^{\pr}+3\mathcal{H}\rho+\partial_{\n}(\rho\, v^{\n})
-\frac{3}{c}\rho\,\varphi^{\pr}+\frac{1}{c^{2}}[(\rho\,v^{2})^\pr
+\partial_{\n}(v^{\n}\rho\, v^{2})+4\mathcal{H}\rho\,v^{2}]
-2\rho\, v^{\n}\partial_{\n}\varphi-\frac{3}{c}\rho\,\psi_{2}^{\pr}
-\frac{6}{c}\rho\,\varphi\varphi^{\pr}=0\,,
\end{align}
while the momentum conservation equation reads
\begin{align}
\label{momentum} 
& (\rho\,v_{\a})^{\pr}+\partial_{\n}\left( \rho\,v^{\n}v_{\a
}\right) +4\mathcal{H} v_{\a}\rho+\rho\,c^{2}\partial_{\a}\varphi
-16\mathcal{H}\rho\,v_{\a}\varphi+\frac{1}{c^{2}}\partial_{\n}\left(
\rho\,v^{2} v^{\n}v_{\a}\right) - 2\rho\,c^{2}\varphi\partial_{\a}
\varphi-\mathcal{H} c\rho\, V_{\a}+\rho\,cv_{\n}\partial_{\a}V^{\n}\nonumber\\
& -4\partial_{\n}\left( \rho\,v^{\n}v_{\a}\varphi\right) +\frac{4}{c^{2}
}\mathcal{H}\rho\,v^{2} v_{\a}-4\rho^{\pr}v_{\a}\varphi-4\rho\,v_{\a}^{\pr
}\varphi-6\rho\,v_{\a}\varphi^{\pr}+\frac{1}{c^{2}}\left( v_{\a}\rho\,
v^{2}\right) ^{\pr}-2\rho\,v_{\a}v^{\n}\partial_{\n}\varphi+2v^{2}
\rho\,\partial_{\a}\varphi+\rho\,c^{2}\partial_{\a}\phi_{2}=0
\end{align}

Eqs.~(\ref{psinew}), (\ref{vnew}) and (\ref{phinew})-(\ref{habnew}), 
together with Eqs.~(\ref{IG0anew}), (\ref{energy}) and (\ref{momentum}), 
are the
main result of our paper and represent a new set of equations which allow to
describe the evolution of metric perturbation from the linear to the strongly
non-linear stage in terms of the gravitational field $\varphi$, the matter
density $\rho$ and the peculiar velocity $v^{\a}$.

\section{Limiting forms of the hybrid approximation in different regimes}

We now show how our approach accounts for known approximation schemes in
different regimes.

\subsection{The linear perturbative regime}

Linearly perturbing Eqs.~(\ref{IG0anew}), (\ref{psinew}) and (\ref{phinew})
with respect to the FRW background we deduce that $\phi$ and $\psi$ coincide
and we obtain the linear scalar potential $\varphi$ in terms of first-order
density and velocity fluctuations,
\begin{align}
\label{Iphipsi}
\nabla^{2} \nabla^{2} \varphi= 4a^{2}\pi G \left[ \nabla^{2}
\delta\rho-3 \mathcal{H}\bar{\rho}\, \partial^{\n}v_{\n}\right] \;,
\end{align}
\begin{align}
\label{linearGoa}
\nabla^{2}(\mathcal{H}\varphi+\varphi^{\pr})= - 4a^{2}\pi G
\bar{\rho}\, \partial^{\n}v_{\n}\;.
\end{align}

Moreover, linearizing Eqs.~(\ref{vnew}),
(\ref{habnew}) and (\ref{trace}), using the linearized expressions for $\delta
T^{i}{}_{j}$ we obtain
\begin{align}
\label{Iphi}
\varphi^{\prime\prime}+3\mathcal{H}\varphi^{\pr}+
(2\mathcal{H}^{\pr}+\mathcal{H}^{2})\varphi=0\;,
\end{align}
\begin{align}
\label{Iv}
\nabla^{2} \nabla^{2} V_{\a}= 0
\end{align}
\begin{align}
\label{Ihab}
h_{\b}^{\prime\prime\alpha} +2\mathcal{H} h_{\b}^{\prime\, \alpha}
-\nabla^{2} h^{\alpha}{}_{\b}=0\;.
\end{align}
Perturbing to first order Eqs.~(\ref{energy}) and (\ref{momentum}), we recover
also the linear continuity and momentum equations which read respectively:
\begin{align}
\label{Ienergy}
\delta\rho^{\pr}+3\mathcal{H}\delta\rho+\bar{\rho}
\,\partial_{\n}v^{\n}-3\bar{\rho}\,\varphi^{\pr}=0\;,
\end{align}
\begin{align}
\label{Imomentum}
v^{\prime\,\alpha}+\mathcal{H} v^{\a}+\partial^{\a}\varphi=0.
\end{align}
In other words, we obtain all the results of linear perturbation theory (see
e.g. Ref.~\cite{mfb,ks}), if we interpret $\varphi$ as the linear scalar
potential.

\subsection{The second-order perturbative regime}

On the other hand, selecting only the growing-mode solution of 
Eq.~(\ref{Iphi}) and perturbing up to second order 
Eqs.~(\ref{psinew})-(\ref{habnew}), in the limit of a pressureless and
irrotational fluid,
we recover all the results of second-order perturbation theory
\cite{mmb,mhm,bruni}. 

More specifically, the first-order vector metric
perturbations vanish, while the linear tensor metric perturbations are
negligible for every kind of cosmological sources, thus we can safely neglect
terms which can be expressed as products of first-order scalar and tensor
metric perturbations. Writing $\varphi(\mathbf{x},\eta) = \varphi
_{0}(\mathbf{x}) g(\eta)$, where $\varphi_{0}$ is the peculiar gravitational
potential linearly extrapolated to the present time and $g \equiv D_{+}/a$ is
the so-called growth-suppression factor, where $D_{+}(\eta)$ is the linear
growing-mode of density fluctuations in the Newtonian limit, and using the
results of the previous subsection we obtain
\begin{align}
\label{IIpsi}
\nabla^{2} \nabla^{2} \psi_{s} & =-4\pi Ga^{2} \left[
3\mathcal{H} \left( \bar{\rho}\,\partial^{\n}v_{s\,\nu} +\delta\rho
\partial^{\n}v_{\n}+ v_{\n}\partial^{\n}\delta\rho- 4\bar{\rho}\,\partial^{\n
}(\varphi v_{\n})\right) -3\bar{\rho}\,\partial^{\n}(\varphi^{\pr}v_{\n
})-\nabla^{2}(\delta\rho_{s}+\bar{\rho}v^{2})\right] \nonumber\\
& -\nabla^{2} \left( \frac{3}{2} \partial^{\n}\varphi\partial_{\n}
\varphi+4\varphi\nabla^{2}\varphi\right) \;,
\end{align}
\begin{align}
\label{IIphi}
\nabla^{2} \nabla^{2}\phi_{s} & =-4\pi Ga^{2} \Bigl[3\mathcal{H}
\left( \bar{\rho}\,\partial^{\n}v_{s\,\nu} +\delta\rho\partial^{\n}v_{\n}+
v_{\n}\partial^{\n}\delta\rho-4\bar{\rho}\,\partial^{\n}(\varphi v_{\n
})\right)  -3\bar{\rho}\,\partial^{\n}(\varphi^{\pr}v_{\n})-\nabla^{2}
(\delta\rho_{s}+2\bar{\rho}\,v^{2})\nonumber\\
& +3\bar{\rho}\,\partial^{\m}\partial_{\n}(v^{\n}v_{\m})\Bigr] +\frac{7}
{2}\nabla^{2} (\partial_{\n}\varphi\partial^{\n}\varphi)-3\partial_{\n
}\partial^{\m}(\partial_{\m}\varphi\partial^{\n}\varphi)\;,
\end{align}
\begin{align}
\label{IIv}
\nabla^{2} \nabla^{2} V_{\a}=16\pi Ga^{2}\partial^{\n}\left( v_{\n
}\partial_{\a}\delta\rho-v_{\a}\partial_{\n}\delta\rho\right) \;,
\end{align}
\begin{align}
\label{IIhab}
\nabla^{2} \nabla^{2} \left(h^{\prime\prime\,\alpha}{}_{\b}
+2\mh h^{\prime\,\alpha}{}_{\b}-\D h^\a{}_\b\right)=16\pi Ga^2
&\bigl[\D\bigl(\D\mr^\a{}_\b-\p^\a\p_\n\mr^\n{}_\b
-\p_\b\p^\n\mr^\a{}_\n \bigr) \nonumber\\
&+\fr{1}{2}\left(\D\p^\m\p_\n\mr^\n{}_\m\,\de^\a{}_\b
+\p^\a\p_\b\p^\m\p_\n\mr^\n{}_\m \right) \bigr]\;, 
\end{align}
where $R^{\a}{}_{\b}$ has the same analytic form of Eq.~(\ref{Rabnew}), that is
\begin{align}
\label{IIRab}
\mr^\a{}_\b=\r\left(v^\a v_\b-\fr{1}{3} v^2 \,\de^\a{}_\b \right)
-\fr{1}{4\pi Ga^2}\left(\p^\a\vp\, \p_\b\vp
-\fr{1}{3}\p^\n\vp\,\p_\n\vp\,\de^\a{}_\b \right)
-\fr{1}{2\pi Ga^2}\left(\vp\,\p^\a\p_\b\vp
-\fr{1}{3}\vp\,\D\vp\,\de^\a{}_\b\right)\;, 
\end{align}
The subscript $s$ indicates quantities evaluated at the second perturbative
order, $v_{s}^{\a}$ is the velocity $dx^{\a}/d\eta$ perturbed at the
second-order and is related to the second-order spatial part 
$v^{\alpha}_{(2)}/a$ of the 4-velocity by the relation 
$v^{\alpha}_{(2)}=v_{s}^{\a}-\varphi v^{\a}$ \footnote{It is worth to 
notice that the traceless tensor $\mr^\a{}_\b$ in Eq.~(\ref{IIRab}) differs
from the correspondent tensor in Eq.~(13) of \cite{mhm}. Besides the global
$4\pi Ga^2$ factor, the terms containing the gravitational potential $\vp$ are 
not the same in the two cases. Actually, what is important is the 
transverse and traceless part of the gravitational wave sources,
and they happen to be the same, since,
as we will show later, our Eq.~(\ref{IIRab}) can be written 
in the form of Eq.~(13) in \cite{mhm} plus other terms which do not contribute
to the relevant component of the source (see Eq.~(\ref{RabPNbis}) and
(\ref{Rab_eff})).}.

We can also find the equations that describe the evolution of $\delta\rho_{s}$
and $v_{s}^{\a}$ by perturbing up to second order 
Eqs.~(\ref{energy})-(\ref{momentum}) and taking the divergence of the 
latter. In this way we
recover also the second-order energy continuity equation
\begin{align}
\label{IIenergy}
\delta\rho_{s}^{\pr}+3\mathcal{H}\delta\rho_{s} 
+\bar{\rho}\partial_{\n}v_{s}^{\n}+\delta\rho\partial_{\n}v^{\n}
+v^{\n}\partial_{\n}\delta\rho+\mathcal{H}\bar{\rho} v^{2}
+\bar{\rho}(v^2)^\prime-3\delta\rho\varphi^{\pr}
-3\bar{\rho}\psi_{s}^{\pr}-6\bar{\rho}\varphi\varphi^{\pr}
-2\bar{\rho}v^{\m}\partial_{\m}\varphi=0\;,
\end{align}
and the divergence of the second-order momentum conservation equation
\begin{align}
\label{IImomentum}
\mathcal{H} \bar{\rho}\,\partial^{\a}v_{s\alpha}+\bar{\rho
}\,\partial^{\a}v_{s\alpha}^{\pr}+\bar{\rho}\,\nabla^{2}\phi_{s}+\partial^{\a}
& \bigl[4 \mathcal{H} \delta\rho v_{\alpha} -4\mathcal{H}\bar{\rho}\,\varphi
v_{\alpha} +\delta\rho^{\pr}v_{\alpha} +\delta\rho v_{\alpha}^{\pr}-4\bar
{\rho}\,\varphi v_{\alpha}^{\pr}-6\bar{\rho}\,\varphi^{\pr}v_{\alpha}
+\delta\rho\partial_{\a}\varphi\nonumber\\
& -2\bar{\rho}\,\varphi\partial_{\a}\varphi+\bar{\rho}\,\partial_{\n}(v^{\n
}v_{\a})\bigr]=0\;.
\end{align}

\subsection{The Newtonian approximation}

From Eqs.~(\ref{weak-field}) and (\ref{psinew}) up to $\mathcal{O}(1/c^{2})$
we deduce
\begin{align}
\label{Poisson}
\nabla^{2}\psi_{1}=\nabla^{2}\phi_{1}=\nabla^{2}\varphi
=\frac{4\pi Ga^{2}}{c^{2}}\delta\rho\;,
\end{align}
and writing $\varphi\equiv\varphi_{N}/c^{2}$, we recover the Poisson equation
$\nabla^{2}\varphi_{N}=4\pi Ga^{2}\delta\rho$, where the subscript $N$ stands
for \textit{Newtonian}.

Analogously, to leading order in $1/c$,  
Eqs.~(\ref{energy}) and (\ref{momentum}) respectively become the usual
continuity and Euler equations of Newtonian cosmology which apply in the limit
of weak fields and non-relativistic velocities \cite{berts}
\begin{align}
\label{continuity}
\rho^{\pr}+3\mathcal{H}\rho+\partial_{\n}(\rho\,v^{\n})=0\;,
\end{align}
\begin{align}
\label{Euler}
v_{\a}^{\pr}+\mathcal{H} v_{\a}+v_{\n}\partial^{\n}v_{\a
}=-\partial_{\a}\varphi_{N}\;.
\end{align}
The latter equation was obtained taking Eq.~(\ref{momentum}) up to
$\mathcal{O}(1/c^{0})$ and inserting Eq.~(\ref{continuity}). 

In the linear limit Eqs.~(\ref{Poisson})-(\ref{Euler}) become
\begin{align}
\label{IPoisson}
\nabla^{2}\varphi=4\pi Ga^{2}\delta\rho\;,
\end{align}
\begin{align}
\label{Icontinuity}
\delta\rho^{\pr}+3\mathcal{H}\delta\rho+\bar{\rho}
\partial_{\n}v^{\n}=0\;,
\end{align}
\begin{align}
\label{IEuler}
v_{\a}^{\pr}+\mathcal{H} v_{\a}=-\partial_{\a}\varphi_{N}\;.
\end{align}
As we can observe, the equations which characterize the linearized Newtonian
theory, differ from the linearized relativistic ones. In particular, while the
momentum conservation Eqs.~(\ref{Imomentum}) and (\ref{IEuler}) are identical,
the linear energy density conservation Eq.~(\ref{Ienergy}) differs from the 
Newtonian one, Eq.~(\ref{Icontinuity}), by the extra term $-3\bar{\rho}
\varphi^{\pr}$ which does not vanish, even for the pure growing-mode
solution of Eq.~(\ref{Iphi}), owing to the presence of a cosmological constant
contribution to the FRW background.

Moreover, Eq.~(\ref{Iphipsi}) represents the linear relativistic 
generalization of the Poisson equation, since
it includes the contribution of the so-called \textit{longitudinal momentum
density} $\varphi_{f}$ ($\partial_{\n}\varphi_{f}=-4a^{2}\pi G \bar{\rho}
v_{\n}$) which acts as a source term for the linear potential $\varphi$. 
Thus, the Poisson gauge gives the \textit{relativistic} cosmological 
generalization of \textit{Newtonian} gravity \cite{berts}.

\subsection{The highly non-linear regime in the PN approximation}

Finally, we consider the case of cosmic structures, in the highly non-linear
regime, whose size is much larger than their Schwarzschild radius (in order to
avoid non-Newtonian terms in the expressions of the sources). 

Our sources can generate vector and tensor metric perturbations by 
mode-mixing in the non-linear regime. In particular, this mechanism applies 
to dark-matter halos around galaxies and galaxy clusters or, more 
specifically, to the highly
condensed substructures by which these halos are characterized.

We obtain the continuity and momentum equations up to $\mathcal{O}(1/c^{2})$,
the equation describing the evolution of the ($0$-$0$) component of the metric
(\ref{gij}) up to $\mathcal{O}(1/c^{4})$, and the equation for the vector
modes $V_{\a}$ up to $\mathcal{O}(1/c^{3})$, i.e. their 1PN approximation.
Moreover, we describe the scalar mode of the ($\alpha$-$\beta$) component of
the metric (\ref{gij}) up to $\mathcal{O}(1/c^{4})$, i.e. we consider its
second post-Newtonian (2PN) approximation, while we obtain the leading-order
terms in powers of $1/c$ for the source of the tensor modes $h^{\a}{}_{\b}$.

Eqs.~(\ref{psinew})-(\ref{habnew}) in this limit become
\begin{align}
\label{psiPPN}
\nabla^{2}\nabla^{2}\psi=\frac{4\pi Ga^{2}}{c^{2}} 
\nabla^{2}\delta\rho+\frac{4\pi Ga^{2}}{c^{4}}\left[ \nabla^{2}(\rho\,v^{2}
)-3\mathcal{H}\partial^{\n}(v_{\n}\rho)\right]  -\frac{1}{c^{4}}\nabla^{2}
\left( \frac{3}{2}\partial^{\n}\varphi_{N}\partial_{\n}\varphi_{N}
+4\varphi_{N}\nabla^{2}\varphi_{N}\right) \;,
\end{align}
\begin{align}
\label{phiPN}
\nabla^{2}\nabla^{2}\phi & =\frac{4\pi Ga^{2}}{c^{2}} 
\nabla^{2}\delta\rho+\frac{4\pi Ga^{2}}{c^{4}}\left[ 2\nabla^{2}(\rho\,v^{2}
)-3\mathcal{H}\partial^{\n}(v_{\n}\rho) -3\partial^{\m}\partial_{\n}
(\rho\,v^{\n}v_{\m})\right] +\frac{7}{2c^{4}}\nabla^{2} (\partial_{\n}
\varphi_{N}\partial^{\n}\varphi_{N})\nonumber\\
& -\frac{3}{c^{4}}\partial_{\n}\partial^{\m}(\partial_{\m}\varphi_{N}
\partial^{\n}\varphi_{N})\;,
\end{align}
\begin{align}
\label{vPN}
\nabla^{2}\nabla^{2} V_{\a}=\frac{16\pi Ga^{2}}{c^{3}}\left[
\partial_{\a}\partial^{\n}(v_{\n}\rho) -\nabla^{2} (v_{\a}\rho)\right] \;,
\end{align}
\begin{align}
\label{habPPN}
\nabla^{2} \nabla^{2} \left(\frac{1}{c^{2}}h^{\prime\prime\,\alpha}{}_{\b}
+\fr{2\mh}{c^2}h^{\prime\,\alpha}{}_{\b}-\D h^\a{}_\b\right)=
\fr{16\pi Ga^2}{c^4}
&\bigl[\D\bigl(\D\mr^\a{}_\b-\p^\a\p_\n\mr^\n{}_\b
-\p_\b\p^\n\mr^\a{}_\n \bigr) \nonumber\\
&+\fr{1}{2}\left(\D\p^\m\p_\n\mr^\n{}_\m\,\de^\a{}_\b
+\p^\a\p_\b\p^\m\p_\n\mr^\n{}_\m \right) \bigr]\;,
\end{align}
where the post-Newtonian limit of the traceless tensor $\mr^\a{}_\b$ is
\begin{align}
\label{RabPN}
\mr^\a{}_\b=\r\left(v^\a v_\b-\fr{1}{3} v^2 \,\de^\a{}_\b \right)
-\fr{1}{4\pi Ga^2}\left(\p^\a\vn\, \p_\b\vn
-\fr{1}{3}\p^\n\vn\,\p_\n\vn\,\de^\a{}_\b \right)
-\fr{1}{2\pi Ga^2}\left(\vn\,\p^\a\p_\b\vn
-\fr{4\pi G a^2}{3}\vn\,\de\r\,\de^\a{}_\b\right)\,.
\end{align}  
Since, in order to compute the metric coefficients up to the PN approximation,
we only need the terms in $\delta T^{i}{}_{j}$ which satisfy the Newtonian
equations of motions, in Eq~(\ref{RabPN}) we have inserted the Poisson
equation. The 1PN extensions of the Newtonian continuity and Euler equations
respectively read
\begin{align}
\label{PNenergy}
\rho^{\pr}+3\mathcal{H}\rho+\partial_{\n}(\rho\, v^{\n}
)+\frac{1}{c^{2}}\left[ (\rho\,v^{2})^\prime
+\partial_{\n}(v^{\n}\rho\, v^{2})
+4\mathcal{H}\rho\, v^{2}
-2\rho\, v^{\n}\partial_{\n}\varphi_{N}\right] =0\,,
\end{align}
\begin{align}
\label{PNmomentum} 
& \rho\,(v_{\a}^{\pr}+\mathcal{H} v_{\a}+v_{\n}\partial
^{\n}v_{\a}+\partial_{\a}\varphi_{N}) +\frac{1}{c^{2}}\biggl[-4\rho^{\pr}
v_{\a}\varphi_{N}-4\rho\,v_{\a}^{\pr}\varphi_{N}-6\rho\,v_{\a}\varphi_{N}
^{\pr}+\left( v_{\a}\rho\, v^{2}\right) ^{\pr}-2\rho\,v_{\a}v^{\n}\partial
_{\n}\varphi_{N} +2v^{2}\rho\,\partial_{\a}\varphi_{N}\nonumber\\
& +\rho\,\partial_{\a}\phi_{PN} -2\rho\, \varphi_{N}\partial_{\a}
\varphi_{N}-\mathcal{H} \rho\, V_{\a} + 
\rho\,v_{\n}\partial_{\a}V_{\n}-16\mathcal{H}
\rho\,v_{\a}\varphi_{N} -4\partial_{\n}\left( \rho\,v^{\n}v_{\a}\varphi
_{N}\right) +\partial_{\n}\left( \rho\,v^{2} v^{\n}v_{\a}\right)
+4\mathcal{H}\rho\,v^{2} v_{\a}\biggr]=0\;,
\end{align}
where $\phi_{PN}$ is given by the 1PN part of Eq.~(\ref{phiPN}). It can be
worth noting that the sources of the metric coefficients involve only
quantities of Newtonian origin, i.e. they do not contain terms defined in
higher-order approximations.

To conclude this subsection, let us stress that all the PN expressions derived
here are \textit{new}, as they are derived in a different gauge than the usual
post-Newtonian \cite{tomita3,tomita4,ch,w}, or synchronous and comoving one
\cite{mt}.

\section{Vector and tensor modes} 

It can be worth to observe that, in the linear limit, Eq.~(\ref{trace})
becomes
\begin{align}
\label{W}
\varphi_N^{\prime\prime}+3\mathcal{H}
\varphi_N^{\pr}+(2\mathcal{H}^{\pr}+\mathcal{H}^{2})\varphi_N=0\;.
\end{align}
This result is extremely important since it implies that the Newtonian
potential $\varphi_N$ and the linear potential $\varphi$ evolve in the same
way with time. Eq.~(\ref{W}) can be also obtained by mixing
together the Newtonian continuity, Euler and Poisson equations perturbed at
first order. This means that, starting from the same initial
conditions, i.e. from the same primordial potential as given e.g. by 
inflation, the two linear potentials $\varphi_{N}/c^2$ and $\varphi$
will assume the same values in each point and at each time. In other
words, Eq.~(\ref{W}) implies that, in the case of first-order matter
perturbations, it is sufficient to use Newtonian gravity on all
scales, provided that we define a ``Newtonian'' linear density
perturbation $\delta\rho_N$ via the Poisson equation applied to the
linear relativistic potential $\varphi$, even if $\delta\rho_{N}$ differs from
the relativistic density $\delta\rho$, as given by Eq.~(\ref{Iphipsi}).
The previous considerations allow to conclude that, for pure growing-mode 
solutions of Eq.~(\ref{W}), in the case of an irrotational and 
pressureless fluid, Eqs.~(\ref{vPN})-(\ref{RabPN}) 
apply to all cosmologically relevant scales, i.e. 
from super-horizon to the smallest ones, even if the density $\rho$, the 
velocity $v^{\a}$ and the potential
$\varphi$ are required to follow the usual Newtonian hydrodynamical equations. 
In the equations that follow, therefore, we will drop the subscript 
$N$ on the various quantities and write 
\begin{align} 
\label{Poissonbis}
\nabla^{2}\varphi = 4\pi Ga^{2} \delta\rho\;,
\end{align}
\begin{align}
\label{continuitybis}
\rho^{\pr}+3\mathcal{H}\rho+\partial_{\n}(\rho\,v^{\n})=0\;,
\end{align}
\begin{align}
\label{Eulerbis}
v_{\a}^{\pr}+\mathcal{H} v_{\a}+v_{\n}\partial^{\n}v_{\a
}=-\partial_{\a}\varphi\;.
\end{align}

Thus, for the vector modes we have
\begin{align}
\label{vPNbis}
\nabla^{2}\nabla^{2} V_{\a}=\frac{16\pi Ga^{2}}{c^{3}}\left[
\partial_{\a}\partial^{\n}(v_{\n}\rho) -\nabla^{2} (v_{\a}\rho)\right] \;,
\end{align}
and, for the tensor modes,   
\begin{align}
\label{habPPNbis}
\nabla^{2} \nabla^{2} \bigg(\frac{1}{c^{2}}h^{\prime\prime\,\alpha}{}_{\b}
&+\fr{2\mh}{c^2}h^{\prime\,\alpha}{}_{\b}-\D h^\a{}_\b\bigg)=
\fr{16\pi Ga^2}{c^4}
\bigl[\D\bigl(\D\mr^\a{}_\b-\p^\a\p_\n\mr^\n{}_\b
-\p_\b\p^\n\mr^\a{}_\n \bigr)\nonumber\\
&+\fr{1}{2}\left(\D\p^\m\p_\n\mr^\n{}_\m\,\de^\a{}_\b
+\p^\a\p_\b\p^\m\p_\n\mr^\n{}_\m \right) \bigr]
=\fr{16\pi Ga^2}{c^4}\D\D\bigg(\mp^\a{}_\n \mp^\m{}_\b 
-\fr{1}{2}\mp^\a{}_\b \mp^\m{}_\n\bigg)\mr^\n{}_\m \;,
\end{align}
where 
\begin{align}
\label{RabPNbis}
\mr^\a{}_\b&=\r\left(v^\a v_\b-\fr{1}{3} v^2 \,\de^\a{}_\b \right)
-\fr{1}{4\pi Ga^2}\left(\p^\a\vp\, \p_\b\vp
-\fr{1}{3}\p^\n\vp\,\p_\n\vp\,\de^\a{}_\b \right) 
-\fr{1}{2\pi Ga^2}\left(\vp\,\p^\a\p_\b\vp
-\fr{4\pi G a^2}{3}\vp\,\de\r\,\de^\a{}_\b\right)\nonumber\\
&=\r\left(v^\a v_\b-\fr{1}{3} v^2 \,\de^\a{}_\b \right)
+\fr{1}{4\pi Ga^2}\left(\p^\a\vp\, \p_\b\vp
-\fr{1}{3}\p^\n\vp\,\p_\n\vp\,\de^\a{}_\b \right)
-\fr{1}{2\pi Ga^2}\left[\p^\a(\vp\p_\b\vp)
-\fr{1}{3}\p^\n(\vp\p_\n\vp)\,\de^\a{}_\b\right]\;, 
\end{align} 
which represent the most important results 
of our paper, since these equations imply that, in the case of matter 
perturbations, the Newtonian description of the sources of vector and 
tensor metric fluctuations can take into account all the effects of the 
relativistic second-order perturbation theory. 

It is important to stress that the third term on the last line of 
Eq.~(\ref{RabPNbis}) does not
contribute to the source of gravitational waves since it vanishes after
applying the projection operation in Eq.~(\ref{Pab}); 
thus we are allowed to drop it and define as effective source of the
gravitational wave $h^\a{}_\b$ the traceless tensor
\begin{align}
\label{Rab_eff}
\mr^\a_{\textrm{eff}\,\b}
=\r\left(v^\a v_\b-\fr{1}{3} v^2 \,\de^\a{}_\b \right)
+\fr{1}{4\pi Ga^2}\left(\p^\a\vp\, \p_\b\vp
-\fr{1}{3}\p^\n\vp\,\p_\n\vp\,\de^\a{}_\b \right)\;. 
\end{align}
Actually from a post-Newtonian point of view, these equations hold true 
also for a pressureless fluid with a vorticity
contribution to the peculiar velocity $v^\a$, but the reader should not be
surprised if curl terms can be produced even by a
pressureless and irrotational perfect fluid. In fact, the curl of the quantity
$\r v^\a$ is still non-vanishing, even if $v^\a$ is derived from a 
scalar potential. 
The solution of the inhomogeneous gravitational-wave 
equation Eq.~(\ref{habPPNbis}) is given in Appendix A. 

\subsection{Comparison with the quadrupole radiation}

We want to show how the gravitational wave source $\mr^\a_{\textrm{eff}\,\b}$ 
in Eq.~(\ref{Rab_eff}) includes the contribution by the 
reduced quadrupole moment \cite{grav} of the matter distribution
expressed via comoving coordinates
\begin{align}
\label{Quadrupole}  
\mathcal{Q}^\a{}_\b=\int d^3\tx\,\r\left( \tx^\a \tx_\b
-\fr{1}{3} \tx^\n \tx_\n\,\de^\a{}_\b\right)\;.
\end{align}
First of all, let us choose the origin of our coordinates $\emph{O}$ inside
the mass-energy distribution described by the stress-energy 
tensor $\de T^i{}_j$.
Let $\mathbf{x}$ be the vector from $\emph{O}$ to the observation point 
$\emph{P}$ and $\mathbf{\tx}$ the vector from $\emph{O}$ to the
volume element $d^3\tx$.
On scales well inside the Hubble horizon, 
the solution of Eq.~(\ref{habPPNbis}), augmented by an outgoing-wave 
boundary condition, is
\begin{align}
\label{hab_retarded}
h^\a{}_\b(\eta,\mathbf{x})=
\fr{4G}{ac^4}\mp^{\a\;\;\m}_{\;\;\n\;\;\;\b}\int d^3\tx 
\fr{(a^3\mr^\n_{\textrm{eff}\,\m})_{ret}}
{\vert\mathbf{x}-\mathbf{\tx}\vert}\;,
\end{align} 
where the transverse-traceless operator is 
$\mp^{\a\;\;\m}_{\;\;\n\;\;\;\b}\equiv\mp^\a{}_\n \mp^\m{}_\b 
-\fr{1}{2}\mp^\a{}_\b \mp^\m{}_\n$, with $\mp^\a{}_\b$ given by Eq.~(\ref{Pab})
and $\mr^\a_{\textrm{eff}\,\b}$ by Eq.~(\ref{Rab_eff}).
The subscript ``ret'' means the quantity is to be evaluated at the retarded
space-time point 
$\left(\eta-\vert\mathbf{x}-\mathbf{\tx}\vert/c,\mathbf{\tx}\right)$.

Our purpose is to evaluate $h^\a{}_\b$ in the wave-zone, that is far outside
the source region: $\vert\mathbf{x}\vert\equiv r\gg\vert\mathbf{\tx}\vert$,
thus we expand the retarded integral Eq.~(\ref{hab_retarded}) in powers of 
$\mathbf{\tx}/r$ and take only the first term of the multipole expansion
\begin{align}
\label{hab_quadrupole}
h^\a{}_\b(\eta,\mathbf{x})=
\fr{4G}{c^4}\fr{1}{ar}\mp^{\a\;\;\m}_{\;\;\n\;\;\;\b}
\left[a^3\int d^3\tx\, \mr^\n_{\textrm{eff}\,\m}\right]_{ret} \;, 
\end{align} 
where for radially travelling waves 
$\mp^\a{}_\b=\de^\a{}_\b-x^\a x_\b/r^2$.
Eq.~(\ref{hab_quadrupole}) expresses the gravitational waves 
$h^\a{}_\b$ in terms of integrals over 
the ``stress distribution'' $\mr^\a_{\textrm{eff}\,\b}$, 
while Eq.~(\ref{Quadrupole}) represents an integral over the source ``energy
distribution''. In order to make the comparison between these
two equations, we need to convert the spatial components
$T^\a{}_\b$ of the stress-energy
tensor in terms of the time components by means of the conservation
equations $T^{j}_{i;j}=0$.
Since $\mathcal{Q}^\a{}_\b$ in Eq.~(\ref{Quadrupole}) is the Newtonian
quadrupole and the dynamics of the tensor source is also Newtonian, we only 
need the continuity and Euler Eq.~(\ref{continuitybis})-(\ref{Eulerbis}), by
which, after some mathematical manipulations, we obtain
\begin{align}
\label{rhov_av^b}
\int d^3\tx \,(\r \,v^\a v_\b)
=\fr{1}{2}\fr{\p}{\p\eta}\int d^3\tx\,\r\left(\tx^\a v_\b+\tx_\b v^\a\right)
+2\mh\int d^3\tx\,\r\left(\tx^\a v_\b+\tx_\b v^\a\right)
+\fr{1}{2}\int d^3\tx\,\r\left(\tx^\a \p_\b\vp+\tx_\b \p^\a\vp\right)\;,
\end{align}
and
\begin{align}
\label{rhox_av^b}
\int d^3\tx\,\r\left(\tx^\a v_\b+\tx_\b v^\a\right)
=\fr{\p}{\p\eta}\int d^3\tx\,(\r\, \tx^\a \tx_\b)
+3\mh\int d^3\tx\,(\r\, \tx^\a \tx_\b) \;,
\end{align}
where we have dropped surface terms at infinity.

By substituting Eq.~(\ref{rhox_av^b}) into Eq.~(\ref{rhov_av^b}), 
we finally find
\begin{align}
\label{deTab}
\int d^3\tx\, (\r\, v^\a v_\b)
&=\fr{1}{2}\fr{\p^2}{\p\eta^2}\int d^3\tx\,(\r\, \tx^\a \tx_\b)
+\fr{3}{2}\fr{\p}{\p\eta}\int d^3\tx\, \mh(\r\, \tx^\a \tx_\b)
+2\mh\int d^3\tx\,\r\left(\tx^\a v_\b+\tx_\b v^\a\right) \nonumber\\
&-\fr{1}{2}\int d^3\tx\,\r\left(\tx^\a \p_\b\vp+\tx_\b \p^\a\vp\right)\;.
\end{align}
After substituting Eq.~(\ref{deTab}) into Eq.~(\ref{Rab_eff}) 
and using again the continuity equation,
in the wave zone the gravitational wave $h^\a{}_\b$, to leading
order in powers of $1/c$ and $\mathbf{\tx}/r$, reads
\begin{align}
\label{hab_final}
h^\a{}_\b&(\eta,\mathbf{x})=
\fr{2G}{ar\,c^4}\mp^{\a\;\;\m}_{\;\;\n\;\;\;\b}
\Bigg\{a^3\Bigg[\fr{\p^2 \mathcal{Q}^\n{}_\m}{\p\eta^2} 
+7\mh\fr{\p\mathcal{Q}^\n{}_\m}{\p\eta}
+\left(3\mh^\pr+12\mh^2\right)\mathcal{Q}^\n{}_\m\nonumber\\
&-\int d^3\tx\,\r\left(\tx^{\n} \p_\m\vp+\tx_\m \p^\n\vp
-\fr{2}{3}\tx^{\sigma} \p_\sigma\vp \,\de^\n{}_\m\right)
+\fr{1}{2\pi Ga^2}\int d^3\tx\,\left(\p^\n\vp\, \p_\m\vp
-\fr{1}{3}\p^\sigma\vp\,\p_\sigma\vp\,\de^\n{}_\m \right)\Bigg]\Bigg\}_{ret}.
\end{align}

Let us observe that the first line of Eq.~(\ref{hab_final}) recovers 
the known expression of the quadrupole radiation 
in the limit of a flat and static Universe \cite{grav,landau}, 
while contributions on the second line
derive from the back-reaction of the gravitational potential $\vp$ which can
act as a source of gravitational waves.
Moreover, on scales much smaller than the Hubble horizon, 
the last two terms on the first line in Eq.~(\ref{hab_final})
can be neglected in comparison to the first one.
In fact, the typical free fall-time of a mass distribution is 
proportional to $\r^{-1/2}$, while the Hubble time goes as $\r_b^{-1/2}$, 
where ``$b$'' stands for background; 
thus, on small scales, where the density contrasts can be very high, 
the characteristic rate of the 
structure collapse is much larger than the expansion rate. 
This allows to drop the contributions proportional to $\mh$ 
in Eq.~(\ref{hab_final}) and recover the results expected 
well inside the horizon.

\section{Concluding remarks}

The main result of this paper is represented by the set of equations
(\ref{psinew}), (\ref{vnew}), (\ref{phinew})-(\ref{Rabnew}) 
and (\ref{Rab_eff}) , expressing metric perturbation in terms 
of the gravitational field $\varphi$, where 
the matter density $\rho$ and the peculiar velocity $v^{\a}$, satisfy
Eqs.~(\ref{IG0anew}), (\ref{energy}) and (\ref{momentum}). 
These equations,
when applied in a cosmological setting characterized by a pressureless and
irrotational fluid and a cosmological constant,
provide a unified description of cosmological perturbations during their
evolution from the linear to the highly non-linear regime.
On large scales, these equations reduce to the equations of the first and 
second-order perturbation theory developed in the
Poisson gauge, while, on very small scales, where the perturbative approach
is no longer applicable, they describe the evolution of cosmological
perturbations by a PN approximation. Indeed, we 
calculate the ($0$-$0$) and ($0$-$\alpha$) components of the metric
(\ref{gij}) up to the 1PN order, the ($\alpha$-$\beta$) scalar-type component
up to the 2PN order, while we find for the ($\alpha$-$\beta$) tensor-type
component the leading-order source terms in powers of $1/c$. 

We also derive the generalization of the standard Euler-Poisson system of 
equations of Newtonian hydrodynamics, that consistently accounts for all the 
effects up to order $1/c^2$.
The curl term and anisotropic stress, that produce vectors and tensor metric 
perturbations, arise already at the second perturbative order and at the 
strongly non-linear level they are dominated by the contribution of the
high-density contrast and the high peculiar velocity typical of 
small-scale structures. 
It can be worth to stress that the quantities which source
vector and tensor modes are of Newtonian origin on all scales, in the sense 
that they involve only terms that satisfy the Newtonian 
Euler-Poisson system. 
This result is of extreme importance in view of a possible numerical 
implementation of our set of equations, as it implies that one can compute 
directly vector and tensor modes starting from the outputs of N-body 
simulations. 

Finally, it should be stressed that our new set of equations has many possible
cosmological applications such as, for example, the  
evaluation of the stochastic gravitational-wave backgrounds produced by 
CDM halos \cite{Q1,Q2,mm2} and substructures within halos. It can be also  
used to improve the estimate of gravitational lensing effects and 
gravity-induced secondary CMB temperature/polarization anisotropies generated
by small-scale structures \cite{pyne,mm,mhm}. 

\acknowledgments{We warmly thank Carlo Baccigalupi and Marco Bruni for helpful 
discussions. Moreover we wish to thank the anonymous referees 
which contributed with
their suggestions to improve the analytic form and 
physical interpretation of our equations.}

\appendix

\section{Solution of the inhomogeneous gravitational-wave equation}

We can formally write Eq.~(\ref{habPPNbis}) as
\begin{align}
\label{habEND}
h_{\beta}^{\prime\prime\alpha}+2\mathcal{H} 
h_{\beta}^{\prime\,\alpha}-\nabla^{2} h_{\beta}^{\;\alpha} =a^{2} \kappa^{2}
\mathcal{S}^{\a}_{\b}\;,
\end{align}
and solve it in the two cases of matter-domination and
$\Lambda$-domination, when the equation of state of the background fluid
is $p=w\rho$ with $w=0,-1$, respectively. In order to obtain these solutions,
we have Fourier-expanded the functions $h^{\alpha}{}_{\beta} (\eta
,\mathbf{x})$ and $\mathcal{S}^{\alpha}{}_{\beta}(\eta,\mathbf{x})$ and
decomposed them in the so-called $\sigma= +,\times$ polarization modes, as
follows
\begin{align}
\label{h}
h^{\alpha}{}_{\beta}(\eta,\mathbf{x})=\int\frac{d^{3}\mathbf{k}}
{(2\pi)^{3}}h^{\alpha}{}_{\beta}(\eta,\mathbf{k}) e^{i\mathbf{k}
\cdot\mathbf{x}}\,,
\end{align}
\begin{align}
\label{hk}
h^{\alpha}{}_{\beta}(\eta,\mathbf{k})=h_{+}(\eta,\mathbf{k})\,
p^{+\,\alpha}{}_{\beta}(\hat{k}) +h_{\times}(\eta,\mathbf{k})\, p^{\times
\,\alpha}{}_{\beta} (\hat{k})\,,
\end{align}
\begin{align}
\label{S}
\mathcal{S}^{\alpha}{}_{\beta}(\eta,\mathbf{x})=\int\frac
{d^{3}\mathbf{k}}{(2\pi)^{3}} \mathcal{S}^{\alpha}{}_{\beta}(\eta,\mathbf{k})
e^{i\mathbf{k}\cdot\mathbf{x}}\,,
\end{align}
\begin{align}
\label{Sk}
\mathcal{S}^{\alpha}{}_{\beta}(\eta,\mathbf{k})= 
\mathcal{S}_{+}(\eta,\mathbf{k})\, p^{+\,\alpha}{}_{\beta}(\hat{k}) 
+\mathcal{S}_{\times
}(\eta,\mathbf{k})\,p^{\times\,\alpha}{}_{\beta} (\hat{k})\,,
\end{align}
where $p^{+\,\alpha}{}_{\beta}(\hat{k})$ and $p^{\times\,\alpha}{}_{\beta}
(\hat{k})$ are the polarization tensors. 

After imposing the initial conditions
\begin{align}
\label{w} 
&  h_{\sigma}(\eta_{eq},\mathbf{k})=h^{\prime}_{\sigma}(\eta
_{eq},\mathbf{k})=0 \qquad\qquad\qquad\qquad\quad\,\text{if}\;\; w=0 \,,\\
&  h_{\sigma}(\eta_{\Lambda},\mathbf{k})=h_{\sigma,\Lambda}(\mathbf{k}),\quad
h^{\prime}_{\sigma}(\eta_{\Lambda},\mathbf{k})=h^{\prime}_{\sigma,\Lambda
}(\mathbf{k}) \qquad\;\text{if}\;\;w=-1\;,
\end{align}
($\eta_{eq}$ is the conformal time at matter-radiation equality, while
$\eta_{\Lambda}$ corresponds to the time when $\bar{\rho}=\rho_{\Lambda}$), we
obtain for $w=0$
\begin{align}
\label{w=0}
h_{\sigma}(\eta,\mathbf{k})=\frac{a_{0}^{2}\kappa^{2}}{\eta_{0}
^{4}}\, k^{3}\left[  \frac{n_{1}(k\eta)}{k\eta}\int_{\eta_{eq}}^{\eta}
\tilde{\eta}^{8}\frac{j_{1}(k\tilde{\eta})}{k\tilde{\eta}} \mathcal{S}
_{\sigma}(\tilde{\eta},\mathbf{k})d\tilde{\eta}- \frac{j_{1}(k\eta)}{k\eta
}\int_{\eta_{eq}}^{\eta}\tilde{\eta}^{8}\frac{n_{1}(k\tilde{\eta})}
{k\tilde{\eta}} \mathcal{S}_{\sigma}(\tilde{\eta},\mathbf{k})d \tilde{\eta
}\right]  \,,
\end{align}
where $\eta_{0}$ represents the present value of the conformal time. For
$w=-1$ we have
\begin{align}
\label{w=-1}
h_{\sigma}(\eta,\mathbf{k})  &  =(\eta k)^{2}j_{1}(\eta k) \left[
h_{\Lambda,\sigma} \frac{2n_{1}(\eta_{\Lambda}k) +\eta_{\Lambda}
n^{\prime}_{1}(\eta_{\Lambda}k)}{\eta_{\Lambda}} 
-\frac{\kappa^{2}}{H^{2}_{0} k}
\int_{\eta_{\Lambda}}^{\eta} \frac{n_{1}(k \tilde{\eta})}{\tilde{\eta}^{2}}
\mathcal{S}_{\sigma}(\tilde{\eta},\mathbf{k})d\tilde{\eta}\right] \nonumber\\
&  -(\eta k)^{2}n_{1}(\eta k)\left[  h_{\Lambda,\sigma} \frac{2j_{1}
(\eta_{\Lambda}k) +\eta_{\Lambda}j^{\prime}_{1}(\eta_{\Lambda}k)}
{\eta_{\Lambda}} -\frac{\kappa^{2}}{H^{2}_{0} k}\int_{\eta_{\Lambda}}^{\eta}
\frac{j_{1}(k \tilde{\eta})}{\tilde{\eta}^{2}} \mathcal{S}_{\sigma}
(\tilde{\eta},\mathbf{k})d\tilde{\eta}\right]  \,,
\end{align}
where $h_{\sigma,\Lambda}(\mathbf{k})$ is obtained from Eq.~(\ref{w=0}) at
$\eta=\eta_{\Lambda}$, while $j_{1}(\eta k)$ and $n_{1}(\eta k)$ are
respectively the spherical Bessel and Neumann functions of order $n=1$.

\end{document}